\documentclass{emulateapj}
\usepackage{graphicx}
\usepackage[none]{hyphenat}
\usepackage{amsmath}
\usepackage{booktabs}
\usepackage{multirow}
\usepackage{txfonts}
\usepackage{enumitem}
\usepackage{amsmath}
\usepackage{algorithmic}
\usepackage{bm}
\usepackage{mathrsfs}
\usepackage{color}

\usepackage[colorlinks,linkcolor=blue,anchorcolor=blue,citecolor=blue]{hyperref}

\begin{document}

\title{Studying Anisotropy of Compressible Magnetohydrodynamic Turbulence by Synchrotron Polarization Intensity}
\author{Ru-Yue Wang\altaffilmark{1,2}, Jian-Fu Zhang\altaffilmark{1,2}, Fu-Yuan Xiang\altaffilmark{1} }
\email{jfzhang@xtu.edu.cn}
\altaffiltext{1}{Department of Physics, Xiangtan University, Xiangtan, Hunan 411105, China}
\altaffiltext{2}{Guizhou Provincial Key Laboratory of Radio Astronomy and Data Processing, Guiyang, Guizhou 550025, China}

\begin{abstract}
\centering
Based on statistical analysis of synchrotron polarization intensity, we study the anisotropic properties of compressible magnetohydrodynamic (MHD) turbulence. 
The second-order normalized structure function, quadrupole ratio modulus and anisotropic coefficient are synergistically used to characterize the anisotropy of the polarization intensity. On the basis of pre-decomposition data cubes, we first explore the anisotropy of the polarization intensity in different turbulence regimes and find that the most significant anisotropy occurs in the sub-Alfv\'enic regime. Using post-decomposition data cubes in this regime, we then study the anisotropy of the polarization intensity from Alfv\'en, slow and fast modes. Statistics of polarization intensity from Alfv\'en and slow modes demonstrate the significant anisotropy while statistics of polarization intensity from fast modes show isotropic structures, which is consistent with the earlier results provided in Cho \& Lazarian ~(\citeyear{2002PhRvL..88x5001C}). As a result, both quadrupole ratio modulus and anisotropic coefficient for polarization intensities can quantitatively recover the anisotropy of underlying compressible MHD turbulence. The synergistic use of the two methods helps enhance the reliability of the magnetic field measurement.
\end{abstract}

\section{Introduction}
Magnetohydrodynamic (MHD) turbulence is ubiquitous in astrophysical environment (see Armstrong et al. ~\citeyear{1995ApJ...443..209A}; Ferri{\`e}re ~\citeyear{2001RvMP...73.1031F}; Elmegreen \& Scalo ~\citeyear{2004ARA&A..42..211E}), and has a profound impact on many key astrophysical processes,
such as formation of stars (Mac Low \& Klessen ~\citeyear{2004RvMP...76..125M}; McKee \& Ostriker ~\citeyear{2007ARA&A..45..565M}), propagation and acceleration of cosmic rays (Yan \& Lazarian ~\citeyear{2008ApJ...673..942Y}), heat conduction (Narayan \& Medvedev ~\citeyear{2001ApJ...562L.129N}), 
and turbulent magnetic reconnection (Lazarian \& Vishniac ~\citeyear{1999ApJ...517..700L}, hereafter LV99). Thus, studying the properties of MHD turbulence is necessary for thorough comprehension of key astrophysical progresses.

MHD turbulence has long been understood as isotropic (Iroshnikov ~\citeyear{1964SvA.....7..566I}; Kraichnan ~\citeyear{1965PhFl....8.1385K}; Shebalin et al. ~\citeyear{1983JPlPh..29..525S}). However, a significant progress was made by Goldreich \& Sridhar ~(\citeyear{1995ApJ...438..763G}, henceforth GS95) in the understanding of the properties of incompressible MHD turbulence, which was predicted to be anisotropic by considering relative motions of eddies parallel and perpendicular to the magnetic field directions. The anisotropies of the turbulence are related to the eddy scales, that is, the smaller the eddy scale is, the more elongated anisotropic structures the turbulence presents. These relations predicted by GS95 have been confirmed numerically (Cho \& Vishniac ~\citeyear{2000ApJ...539..273C}; Maron \& Goldreich~\citeyear{Maron01}; Cho \& Lazarian ~\citeyear{2002PhRvL..88x5001C}, henceafter CL02; Cho \& Lazarian ~\citeyear{2003MNRAS.345..325C}, henceafter CL03). It is worth pointing out that the GS95 theory was applicable for the local system of reference defined by the direction of the local mean magnetic field surrounding the eddies (LV99; Cho \& Vishniac ~\citeyear{2000ApJ...539..273C}; Cho et al.~\citeyear{cho2002simulations}). This definition provides an important basis for understanding of the critical balance condition, the scale-dependent anisotropy and turbulent reconnection theory. The GS95 theory focused on incompressible MHD turbulence which was extended to compressible ones (CL02 and CL03). The anisotropies of compressible MHD turbulence were explored in the subsequent numerical simulation (Matthaeus et al.~\citeyear{1996JGR...101.7619M}), which opened the way for numerical simulation of the compressibility of MHD turbulence. An important work of compressible MHD turbulence was carried out by CL02 where they developed and tested the technique of separating different MHD modes. By decomposing the MHD turbulence modes, they found that Alfv\'en and slow modes have scale-dependent GS95 anisotropy while fast mode is isotropic.

In fact, the compressible MHD turbulence is involved in density and magnetic field fluctuations. The plasma parameter $\beta=P_{\rm gas}/P_{\rm mag}$ is used to characterize the compressibility of MHD turbulence, where $P_{\rm gas}$ and $P_{\rm mag}$ are gas and magnetic pressures, respectively.  By introducing the Alfv\'enic and sonic Mach numbers:
\begin{equation}
M_{\rm A}=\langle \frac{\vert {\bm v}\vert}{v_{\rm A}}\rangle,~~and~~
M_{\rm s}=\langle \frac{\vert {\bm v}\vert}{c_{\rm s}}\rangle, \label{eq:1}
\end{equation}
respectively, we have $\beta=2M_{\rm A}^2/M_{\rm s}^2$. Here, ${\bm v}$ is the gas velocity, $v_{\rm A}=\vert {\bm B}\vert/\sqrt{4\pi \rho}$ is Alfv\'enic velocity and $c_{\rm s}$ is sound speed. High-$\beta$ regime indicates a gas-pressure-dominated plasma, whereas low-$\beta$ regime a magnetic-pressure-dominated one. The incompressible regime formally corresponds to $\beta\sim \infty$, and the compressible one to low-$\beta$ regime. 

Although numerical simulation is a great tool for exploring the properties of MHD turbulence, the currently available 3D simulations are limited by numerical resolutions, with the Reynolds number up to $\sim 10^5$ less than the Reynolds number at least $10^8$ of real astrophysical fluids. Therefore, the interstellar medium (ISM) turbulence measurements are important for studying properties of ISM, gauging the Cosmic Microwave Background (CMB) foregrounds and predicting the propagation of cosmic rays. Obviously, to develop an effective method based on observations is very promising for exploring the properties of turbulent fluctuations.

Based on an analytical description, Lazarian \& Pogosyan~(\citeyear{2012ApJ...747....5L}, henceforth LP12) made the first step into employing statistics of synchrotron intensity fluctuations to recover the properties of MHD turbulence. They predicted that (1) synchrotron intensity fluctuations are anisotropic with the elongated structure along the direction of magnetic field; (2) anisotropies of synchrotron fluctuations are dominated by the quadrupole component, and the quadrupole ratio, i.e., the ratio between quadrupole and monopole parts, is sensitive to the compressibility of underlying turbulence; (3) synchrotron intensity correlations are very weakly dependent on spectral index of relativistic electrons. These theoretical predictions were tested and verified numerically by Herron et al.~(\citeyear{Herron16}) using MHD turbulence simulations. In addition, LP12 also found that the quadrupole ratio moduli of synchrotron intensity from three (Alfv\'en, slow and fast) modes show different levels of anisotropy related to Alfv\'enic Mach numbers, which should be tested further. The advantage of using synchrotron intensity fluctuations to study MHD turbulence is free from the influence of Faraday rotation. However, we all know that this technique is not applicable to the line-of-sight magnetic field measurements. 

In view of the powerful and abundant polarization observations, the traditional Faraday rotation synthesis or Faraday tomography (Burn \citeyear{1966MNRAS.133...67B}; Brentjens \& de Bruyn \citeyear{2005A&A...441.1217B}) has been popularly used in the studies of the properties of magnetic field by combining synchrotron polarization with Faraday rotation effect. This method recently applied to polarized observational data provides a significant insight into magnetic field structures of galaxies (see Fletcher et al. \citeyear{2011MNRAS.412.2396F}; Beck \& Wielebinski \citeyear{2013pss5.book..641B}; Haverkorn \citeyear{2015ASSL..407..483H} for a review) and the properties of the ISM (Jeli\'c et al. \citeyear{2015A&A...583A.137J}; Van Eck et al. \citeyear{2017A&A...597A..98V}; Dickey et al. \citeyear{2019ApJ...871..106D}). An apparent limitation of the Faraday tomography is related to its treatment of magnetic turbulence. In particular, the estimation of the line-of-sight magnetic field through Faraday depth is ambiguous due to magnetic field reversals along the line of sight. With the purpose of measuring the properties of MHD turbulence, statistical techniques of synchrotron polarization gradient, i.e., the gradient of polarized vector, have been applied to determining the sonic Mach number of the interstellar turbulence (Gaensler et al. \citeyear{2011Natur.478..214G}; Burkhart et al. \citeyear{2012ApJ...749..145B}). Very recently, more sophisticated constructions of synchrotron diagnostics have been derived in Herron et al.~(\citeyear{2018bApJ...853....9H}) for feasibly exploring magnetized ISM.

Taking synchrotron polarization fluctuations into account, Lazarian \& Pogosyan~(\citeyear{2016ApJ...818..178L}, henceforth LP16) introduced two important new techniques, i.e., polarization spatial analysis and polarization frequency analysis, to obtain the properties of magnetic turbulence, such as the power spectral index and correlation scale. These two techniques consider polarization fluctuations as a function of the spatial separation of the direction of the measurements and wavelength for the same line of sight, respectively. The theoretical predictions presented by LP16 have been successfully tested by Zhang et al.~(\citeyear{2016ApJ...825..154Z,2018ApJ...863..197Z}). With the modern understanding of MHD turbulence theory (GS95) and analytical theoretical description of synchrotron intensities (LP12) and polarization intensities (LP16), Lazarian \& Yuen~(\citeyear{2018bApJ...865...59L}) developed a new (scalar quantity) gradient technique of synchrotron polarization intensities, different from the gradient of polarization vector provided by Gaensler et al. (\citeyear{2011Natur.478..214G}), for tracing the direction of mean magnetic fields. New gradient technique originates from the relation between the magnetic field direction and synchrotron gradients according to the modern understanding of fundamental theory of magnetic turbulence, i.e., the eddies aligning with the local magnetic field surrounding the eddies. This technique has been generalized from sub-Alfv\'enic  to super-Alfv\'enic turbulence regime in the terms of the multi-frequency measurement (Zhang et al.~\citeyear{2019MNRAS.486.4813Z}). By analogy to anisotropies for synchrotron intensity fluctuations (LP12), LP16 predicted that polarization intensity fluctuations should show similar behavior recovering the underlying magnetic field fluctuations.

It is obvious that not only can recovering anisotropy of MHD turbulence explore the structure of the eddies but also trace the direction of mean magnetic field. Besides, studying compressibility allows us to understand better relations between turbulent magnetic filed and densities. As is mentioned above, a large number of the properties of compressible MHD turbulence are derived from direct numerical simulation. But the current numerical capabilities are not sufficient to simulate a real astrophysical environment, we do not know to what extent the properties and laws based on finite numerical simulations are correct. This motivates us to explore the properties of compressible MHD turbulence using mimic polarization observations. We expect to know whether statistical analysis techniques of polarization intensities can reveal the properties of both anisotropy and compressibility of MHD turbulence. 

The structure of the paper is organized as follows. In Section \ref{section2}, we provide theoretical descriptions including MHD turbulence basics,
synchrotron radiative processes and measurement techniques of synchrotron polarization fluctuations. Section \ref{section3} presents the procedure of numerical
simulation of MHD turbulence. Anisotropies of polarization intensity are presented in Section \ref{aniTotMod} for the mode pre-decomposition scenario and in Section \ref{aniThreeMod} for the mode post-decomposition case. Discussion and a brief summary are given in Sections \ref{Discus} and \ref{Summ}, respectively.

%%%%%%%%%%%%%%%%%%%%%%%%%%%%%%%%%%%%%%%%%%%%%%%%%%%%%%
%%                                      Section 2                                       %%%%%%%%%%%%%%%%%%%%%%%
%%%%%%%%%%%%%%%%%%%%%%%%%%%%%%%%%%%%%%%%%%%%%%%%%%%%%%

\section{Theoretical Descriptions}\label{section2}
\subsection{Fundamental of MHD Turbulence}\label{section2.1}
The modern MHD turbulence theory states a collection of anisotropic eddies in the presence of strong magnetic field (GS95), the direction of which is aligned with the major axis of the eddies. In general, this alignment law can be remained only when the energy of magnetic field over the volume of the eddy is equivalent to
or larger than the kinetic energy of the eddy. The Alfv\'enic Mach number mentioned above can quantificationally characterize this relation. This parameter can be used to describe different MHD regimes. The incompressible MHD turbulence by GS95 corresponds to $M_{\rm A}\sim1$ as an example. Under the condition of a critical balance, i.e., $v_{l}l_{\perp}^{-1}=V_{\rm A}l_{\|}^{-1}$, where $v_ l$ is the fluctuation velocity at the scale $l$, with $l_{\|}$ and $l_{\perp}$ indicating parallel and perpendicular scales of the eddies respectively, the derived relation of 
\begin{equation}
l_{\|}\propto l_{\perp}^{2/3} \label{eq:aniso}
\end{equation}
characterizes the anisotropy of MHD turbulence.

The generalization of the GS95 theory to $M_{\rm A}<1$ and $M_{\rm A}>1$ was provided in LV99 and Lazarian~(\citeyear{2006ApJ...645L..25L}).  In the case of the $M_{\rm A}<1$, when the turbulence is driven with sub-Alfv{\'e}nic velocities at injection scale $L_{\rm inj}$, there is a range of weak turbulence that spans from $L_{\rm inj}$ to the transition scale $l_{\rm trans}=L_{\rm inj}M_{\rm A}^2$. The strong sub-Alfv{\'e}nic turbulence is present on the scales less than $l_{\rm trans}$. The eddies of magnetic turbulence are more elongated along the magnetic field compared to the original GS95 anisotropy (see Equation \ref{eq:aniso} ). Over the inertial range of $[l_{\rm trans}, l_{\rm diss}]$, where $l_{\rm diss}$ is the turbulence dissipation scale, the relation between the extended eddy scale along the magnetic field $l_\|$ and the transversal eddy scale $l_\bot$ is  
\begin{equation}
l_{\|}\approx L_{\rm inj}^{1/3}l_{\perp}^{2/3}M_{\rm A}^{-4/3}, \label{anis}
\end{equation}
when $M_A=1$ returns to the original GS95 relation. What can be obtained in the case of $M_{\rm A}>1$ corresponds to super-Alfv{\'e}nic turbulence $V_{\rm L}> {V}_{{\rm{A}}}$. For a limiting case of ${M}_{{\rm{A}}}\gg 1$, since the weak magnetic field has a marginal influence on MHD turbulence, the turbulence at scale close to $L_{\rm inj}$ scale has an essentially hydrodynamic Kolmogorov property, i.e., $v_l=V_{\rm L}(l/L_{\rm inj})^{1/3}$. The hydrodynamic properties of cascade processes have been changed at the scale $l_{\rm A}=L_{\rm inj}M_{\rm A}^{-3}$, where the turbulent velocity is equal to the Alfv{\'e}n velocity $v_l=V_{\rm A}$ (Lazarian \citeyear{2006ApJ...645L..25L}). In the inertial range of $[l_{\rm A}, l_{\rm diss}]$, the super-Alfv\'enic turbulence shows again the characteristics of the GS95 anisotropy.
 
The GS95 theory developed in the study of incompressible turbulence serves as a guide in the later exploration of compressible MHD turbulence. The applicability of the model of real compressible turbulence was obtained in numerical simulations (Lithwick \& Goldreich~\citeyear{Lithwick01}, CL02 and CL03). Specifically, CL02 and CL03 presented a technique of decomposing MHD turbulence into three modes, i.e., Alfv\'en, slow and fast modes, which made a great contribution to the development of MHD turbulence. They demonstrated that while the density is greatly modified as a result of the compressibility, the magnetic and velocity fluctuations of Alfv{\'e}n and slow modes are only marginally different from the incompressible case. Meanwhile, numerical simulations show that the amount of energy in fast modes is less than that in Alfv{\'e}n and slow modes (CL02, and Kowal \& Lazarian~\citeyear{2010ApJ...720..742K}). Among them, Alfv{\'e}n modes follow Kolmogorov spectrum ($l_{\|} \propto l_{\perp}^{2/3}$) and are compatible with the GS95 anisotropic model, while slow modes ($l_{\|} \propto l_{\perp}^{2/3}$) sheared by Alfv\'en modes are likely to evolve passively and fast modes have isotropic properties ($l_{\|} \propto l_{\perp}$). 

\subsection{Synchrotron Radiative Processes}\label{section2.2}
The relativistic electrons merged into the magnetic field are accelerated by the Lorentz force, resulting in the production of synchrotron radiation. Since synchrotron fluctuations carry the information of turbulent magnetic fluctuations, developing a statistical technique related to synchrotron observations is a natural way for the study of MHD turbulence. We assume a homogeneous and isotropic power-law energy distribution of relativistic electrons 
\begin{equation}
N(E) dE=KE^{2\alpha-1}dE, \label{eq:2}
\end{equation}
where $N$ is the number density of relativistic electrons with the energy between $E$ and $E+dE$, $K$ is a normalization constant, and $\alpha$ is a spectral index of the electrons. 

The synchrotron emission intensity as a function of radiative frequency $\nu$ is given by (Ginzburg \& Syrovatskii~\citeyear{1965ARA&A...3..297G})
\begin{eqnarray}
I(\nu)=&&  \frac{e^3}{4\pi m_{\rm e} c^2} \int_0^L \frac{\sqrt 3}{2-2\alpha}
\Gamma \left(\frac{2-6\alpha}{12}\right)\Gamma \left(\frac{22-6\alpha}{12}\right)
\nonumber \\
&&  \times \left(\frac{3e}{2\pi m_{\rm e}^3 c^5}\right)^{-\alpha} 
KB_{\perp}^{1-\alpha} \nu^{\alpha}dL, 
\label{eq:I}
\end{eqnarray}
where $\Gamma$ indicates the gamma function, $B_{\perp}$ represents the strength of magnetic field perpendicular to the line of sight and $L$ is an integral length along the line of sight. Other parameters ($e$, $m_{\rm e}$ and $c$) keep their usual meanings. 

By means of the synchrotron intensity provided in Equation (\ref{eq:I}) and the fraction polarization degree ($p$), the linearly polarized intensity can be calculated by 
\begin{equation}
PI=I\times p=I\times \frac{3-3\alpha}{5-3\alpha}. \label{eq:PI1}
\end{equation}
The Stokes parameters $Q$ and $U$ are related to the polarization intensity by $Q=PI\cos 2\psi$ and $U=PI\sin 2\psi$. From the view point of an observation, the polarization intensity $PI$ and the polarization angle $\psi$ (measured anti-clockwise from North) are intuitive measurement for the linearly polarized emission (Hamaker $\&$ Bregman~\citeyear{1996A&AS..117..161H}) and are given by 
\begin{equation}
PI=\sqrt{Q^2+U^2} ~~and ~~\psi=\frac{1}{2}\arctan{\frac{U}{Q}}. \label{eq:PI2}
\end{equation}
Besides, we can define the complex polarization vector as $\bm P=Q+iU$, by which the polarization intensity is written as the complex modulus of $\vert \bm P \vert$ and complex argument as $2\psi$ in the $Q$-$U$ plane. With the consideration of Faraday rotation effect, the polarization angle is expressed as $\psi=\psi_{0}+{\rm RM} \lambda^2$, where $\lambda$ is the wavelength of synchrotron radiation and the rotation measure is given by ${\rm RM}=0.81\int_0^L n_{\rm e}B_{\parallel}\,dz~\rm rad~m^{-2}$, where $n_{\rm e}$ is the number density of thermal electrons and $B_{\parallel}$ is the strength of magnetic field parallel to the line of sight.

\subsection{Statistical Measure Techniques}\label{section2.3}
To extract the properties of MHD turbulence, basic statistical tools, such as the power spectrum, power density spectrum, correlation function and structure function, are adopted. In this paper, both the structure function and a new quadrupole ratio for synchrotron polarization intensities are used to explore the anisotropy of compressible MHD turbulence. 

Based on LP12, the normalized correlation function (NCF) of synchrotron polarization intensity is written as 
\begin{equation}
\xi_{\rm PI}(\bm R)=\frac
{{\langle PI(\bm X)PI(\bm{X+R})\rangle}-{\langle PI(\bm X)\rangle}^2}
{\langle PI(\bm X)^2 \rangle-{\langle PI(\bm X)\rangle}^2},\label{eq:NCF}
\end{equation}
where $\bm{X}=(x,y)$ denotes a two-dimensional position vector, and $\bm R$ is a separation vector between any two points on the plane of the sky. By analogy of the formula of LP12, the structure function of synchrotron polarization intensity is given by 
\begin{equation}
D_{\rm PI}(\bm R)=\langle(PI(\bm X)-PI(\bm{X+R}))^2\rangle.\label{eq:7}
\end{equation}
Alternatively, the normalized structure function of polarization intensity can be obtained by 
\begin{equation}
\tilde{D}_{\rm PI}=2(1-\xi_{\rm PI}), \label{eq:NSF}
\end{equation}
 according to Equation (\ref{eq:NCF}). By Equation (\ref{eq:NSF}), the quadrupole ratio for synchrotron polarization intensities is defined by LP12
\begin{equation}
\frac{\tilde{M}_{\rm 2}(R)}{\tilde{M}_{\rm 0}(R)}=
\frac{\int_0^{2\pi} e^{-2i\phi} \tilde{D}_{\rm PI}(R,\phi)~d{\phi}}
{\int_0^{2\pi} \tilde{D}_{\rm PI}(R,\phi)~d{\phi}}, \label{eq:9}
\end{equation}
where $R$ is a radial separation between two points, and $\phi$ represents the polar angle. The quadrupole ratio is used to reveal the anisotropy of MHD turbulence.

In addition, we can employ a diagnostic of the anisotropic coefficient to quantitatively characterize the anisotropy of the turbulence. Specifically, we consider two measurement directions perpendicular to each other from polarization intensities and define their ratio of individual structure function,
\begin{equation}
AC=\frac{{SF}_{\parallel}}{SF_{\perp}}=
\frac{\langle{\vert PI_{\parallel}(x+r_{\parallel})-PI_{\parallel}(x)\vert}^2\rangle}
{\langle{\vert PI_{\perp}(y+r_{\perp})-PI_{\perp}(y)\vert}^2\rangle}, \label{eq:AC}
\end{equation}
as an anisotropic coefficient in dimensionless units, with the relation of $r=[r_{\parallel}^2+r_{\perp}^2]^{1/2}$. Equation (\ref{eq:AC}) indicates that the isotropy and anisotropy of the turbulence structure happened respectively in the case of $AC=1$ and that of $AC$ far away from 1.
\begin{deluxetable*}{cccccc}

\tabletypesize{\scriptsize}
%\rotate
\tablecaption{Data cubes with numerical resolution of $512^3$ generated in the simulation of compressible MHD turbulence. $\delta B_{\rm rms}$ indicates the root mean square of random magnetic field and ${\langle B \rangle}$ the regular magnetic field. }
\tablewidth{170mm}
\tablehead{
\colhead{run} & \colhead{$M_{\rm s}$} 
   & \colhead{$M_{\rm A}$}
   & \colhead{$\beta$}
   & \colhead{$\delta B_{\rm rms}/{\langle B \rangle}$}
   & \colhead{Turbulence regime}
}
\startdata
1   & 9.92   & 0.50   &0.005    &0.465   & Supersonic and sub-Alfv\'enic \\
2   & 6.78   & 0.52   &0.012    &0.463   & Supersonic and sub-Alfv\'enic \\
3   & 4.46   & 0.55   &0.030  &0.467   & Supersonic and sub-Alfv\'enic \\
4   & 3.16   & 0.58   &0.067    &0.506   & Supersonic and sub-Alfv\'enic \\
5   & 0.87   & 0.70   &1.295    &0.579   & Subsonic and sub-Alfv\'enic \\
6   & 0.48   & 0.65   &3.668    &0.614   & Subsonic and sub-Alfv\'enic \\
7   & 7.02   & 1.76   &0.126    &4.606   & Supersonic and super-Alfv\'enic \\
8   & 4.32   & 1.51   &0.244    &5.175   & Supersonic and super-Alfv\'enic \\
9   & 3.11   & 1.69   &0.591    &5.254   & Supersonic and super-Alfv\'enic \\
10  & 0.83   & 1.74   &8.790    &6.110   & Subsonic and super-Alfv\'enic \\
11  & 0.45   & 1.72   &29.219   &6.345   & Subsonic and super-Alfv\'enic  
\enddata
\label{table_1}
\end{deluxetable*}

%%%%%%%%%%%%%%%%%%%%%%%%%%%%%%%%%%%%%%%%%%%%%%%%%%%%%%
%%                                      Section 3                                       %%%%%%%%%%%%%%%%%%%%%%%
%%%%%%%%%%%%%%%%%%%%%%%%%%%%%%%%%%%%%%%%%%%%%%%%%%%%%%
\section{Simulation methods and data generation}\label{section3}
The third-order-accurate hybrid, essentially non-oscillatory code, is used to solve control equations of MHD turbulence as follows 
\begin{equation}
{\partial \rho }/{\partial t} + \nabla \cdot (\rho {\bm v})=0, \label{eq:11}
\end{equation}
\begin{equation}
\rho[\partial {\bm v} /{\partial t} + ({\bm v}\cdot \nabla) {\bm v}] +  \nabla p
        - {\bm J} \times {\bm B}/4\pi ={\bm f}, \label{eq:12}
\end{equation}
\begin{equation}
{\partial {\bm B}}/{\partial t} -\nabla \times ({\bm v} \times{\bm B})=0,\label{eq:13}
\end{equation}
\begin{equation}
\nabla \cdot {\bm B}=0,\label{eq:14}
\end{equation}
and simulate an emitting synchrotron radiation medium in a periodic box of the length of $2\pi$. Here, $p=c_{\rm s}^2\rho$ is gas pressure, $t$ is the evolution time of the fluid, ${\bm J}=\nabla \times {\bm B}$ is the current density, and ${\bm f}$ is a random driving force. In the simulation, the turbulence is driven by a random solenoidal driving force on large scales and an isothermal equation of state is used to close the above equations. Furthermore, non-zero mean magnetic fields are set along the $x$-axis direction. The obtained data cubes with numerical resolution of $512^3$, corresponding to different turbulence regimes, are listed in Table \ref{table_1}. These data cubes are mainly characterized by the Alfv\'enic and sonic Mach numbers. 

The decomposition method of MHD modes provided by CL02 is used to decompose data cubes listed in Table \ref{table_1} into Alfv\'en, slow and fast modes. \footnote{Here, each MHD mode is separated by the Fourier transformation related to the mean magnetic field. Since affected by the wandering of large scale magnetic field and density inhomogeneities, this approach is applicable for the case of the sub-Alfv{\'e}nic turbulence with a strong mean magnetic field and small perturbation. An improved procedure is to extend the CL03 decomposition into modes, in which each components of the local magnetic field are decomposed into orthogonal wavelets using discrete wavelet transformation (Kowal \& Lazarian ~\citeyear{2010ApJ...720..742K}). This method has more significant advantages in dealing with turbulent fluctuations of high amplitude such as trans-Alfv\'enic turbulence.} The main procedures are given as follows:
\begin{equation}
\hat{\zeta}_{\rm f} \varpropto (1+\frac{\beta}{2} + \sqrt{D})(k_{\perp} \hat {\bm k}_{\perp}) 
+(-1 + \frac{\beta}{2}+ \sqrt{D})(k_{\parallel} \hat {\bm k}_{\parallel}), \label{eq:15}
\end{equation}
\begin{equation}
\hat{\zeta}_{\rm s} \varpropto (1+\frac{\beta}{2} - \sqrt{D})(k_{\perp} \hat {\bm k}_{\perp}) 
+(-1 + \frac{\beta}{2}- \sqrt{D})(k_{\parallel} \hat {\bm k}_{\parallel}), \label{eq:16}
\end{equation}
\begin{equation}
\hat{\zeta}_{\rm A} \varpropto -\hat{\bm k}_{\perp} \times \hat{\bm k}_{\parallel},\label{eq:17}
\end{equation}
where $D=(1+\frac{\beta}{2})^2-2 \beta \cos^2\theta$ and $\cos\theta=\hat{k}_{\parallel} \cdot \hat{B}$. 
We only use the line-of-sight component of decomposed velocities to calculate the magnetic field mode components, resulting in three magnetic field modes
\begin{equation}
B_{\rm (f,s,a),z}=[\mathcal{F}^{-1}(\mathcal{F}(\bm B)\cdot \hat{\zeta}_{\rm f,s,a})]
(\hat{\zeta}_{\rm f,s,a} \cdot \hat{\zeta}_{\rm LOS}), \label{eq:18}
\end{equation}
where $\mathcal{F}$ is the Fourier transform operator.

Those data cubes decomposed by the method of CL02 are employed as input parameters of Equations (\ref{eq:I}) and (\ref{eq:PI1}) for reprocessing simulation to construct a synthetic map of synchrotron emission. Assuming the line of sight ($z$-axis) perpendicular to the direction of mean magnetic fields ($x$-axis), we can calculate $B_{\perp}$ in Equation (\ref{eq:I}) by $B_{\perp}=\sqrt{B_{\rm x}^2+B_{\rm y}^2}$. The final synthetic synchrotron map is generated by the line-of-sight integral to simulate real observations.

%%%%%%%%%%%%%%%%%%%%%%%%%%%%%%%%%%%%%%%%%%%%%%%%%%%%%%
%%                                      Section 4                                       %%%%%%%%%%%%%%%%%%%%%%%
%%%%%%%%%%%%%%%%%%%%%%%%%%%%%%%%%%%%%%%%%%%%%%%%%%%%%%

\section{Anisotropy of polarization intensity for pre-decomposition MHD modes}\label{aniTotMod}
Before decomposing compressible MHD turbulence modes, we explore in this section anisotropy of pre-decomposition MHD modes by statistics of synchrotron polarized intensity. Using different statistical techniques and data cubes listed in Table \ref{table_1}, we investigate how Mach numbers (Section \ref{EffMaMs1}) and radiation frequency (Section \ref{EffFreq1}) affect anisotropy of polarized intensity reflecting the anisotropy of MHD turbulence.   

\subsection{Effect of Mach Numbers}\label{EffMaMs1}

\begin{figure*}[ht]
\center
\includegraphics[width=0.80\textwidth]{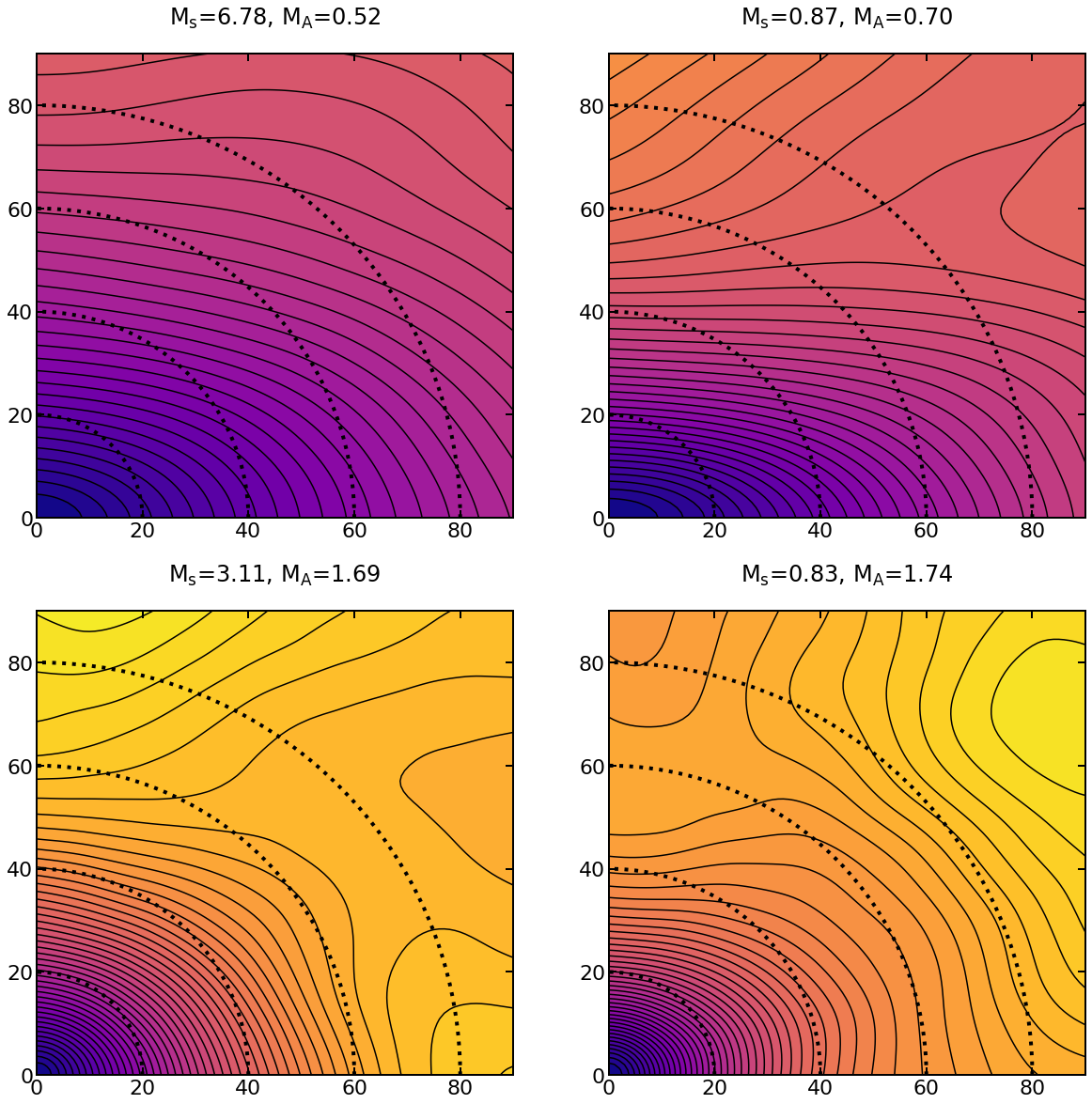}
\caption{Images of the normalized structure function of synchrotron polarization intensities at different turbulence regimes. The dotted contour lines plotted in each panel denote isotropy.
}
\label{Fig:ConDiffMaMs}
\end{figure*}

We employ Equation (\ref{eq:PI1}) to obtain synthetic map of synchrotron polarization intensity with the setting of $\alpha=-1$ and $\nu=10$ GHz, corresponding to the weak Faraday rotation effect.  It is known that structure function is the most intuitive method to reflect structure anisotropy. Here, we focus mainly on using  the second-order normalized structure function (see Equation \ref{eq:NSF}) to reveal the anisotropy of polarization intensity at different turbulence regimes. The resulting contour maps of normalized structure function of polarization intensity are shown in Figure \ref{Fig:ConDiffMaMs}, at supersonic and sub-Alfv\'enic (left upper panel: $M_{\rm s}=6.78$, $M_{\rm A}=0.52$ ), subsonic and sub-Alfv\'enic (right upper panel: $M_{\rm s}=0.87$, $M_{\rm A}=0.70$), supersonic and super-Alfv\'enic (left lower panel: $M_{\rm s}=3.11$, $M_{\rm A}=1.69$), as well as subsonic and super-Alfv\'enic (right lower panel: $M_{\rm s}=0.83$, $M_{\rm A}=1.74$) regimes. The pixels of the map is in units of the code unit, and the dotted contour lines are plotted to indicate isotropy for the sake of comparison. 

As is shown in Figure \ref{Fig:ConDiffMaMs}, the solid contour lines are extended along the $x$-axis (horizontal) directions, which reflects the anisotropic structure of eddies whose major axis is aligned with the direction of mean magnetic field. The right upper panel presents more elongated structure of normalized polarization intensity than that of the left upper panel, however, anisotropic structures in the right upper panel are destroyed on a large scale. Similarly, on smaller scale ($<60$ pixels) than those of the upper panels, the right lower panel also presents slightly more extended structure than that of the left lower panel, both of which show the structure destroyed on a large scale. As a result, the anisotropy of synchrotron polarization intensity is more sensitive to sub-Alfv\'enic turbulence than to super-Alfv\'enic one. The reason is that sub-Alfv\'enic turbulence corresponds to stronger magnetic field fluctuation and produces more significant anisotropy. Although structures of polarized intensity have weak dependence on sonic Mach numbers, it can been seen that subsonic turbulence produces more anisotropic structure at small scales than supersonic one. In addition, it may be that supersonic Mach number helps to form the structure of anisotropy on larger scales (see left upper panel) due to the formation of shock waves.

\begin{figure*}[ht]
\centering
\includegraphics[width=0.80\textwidth]{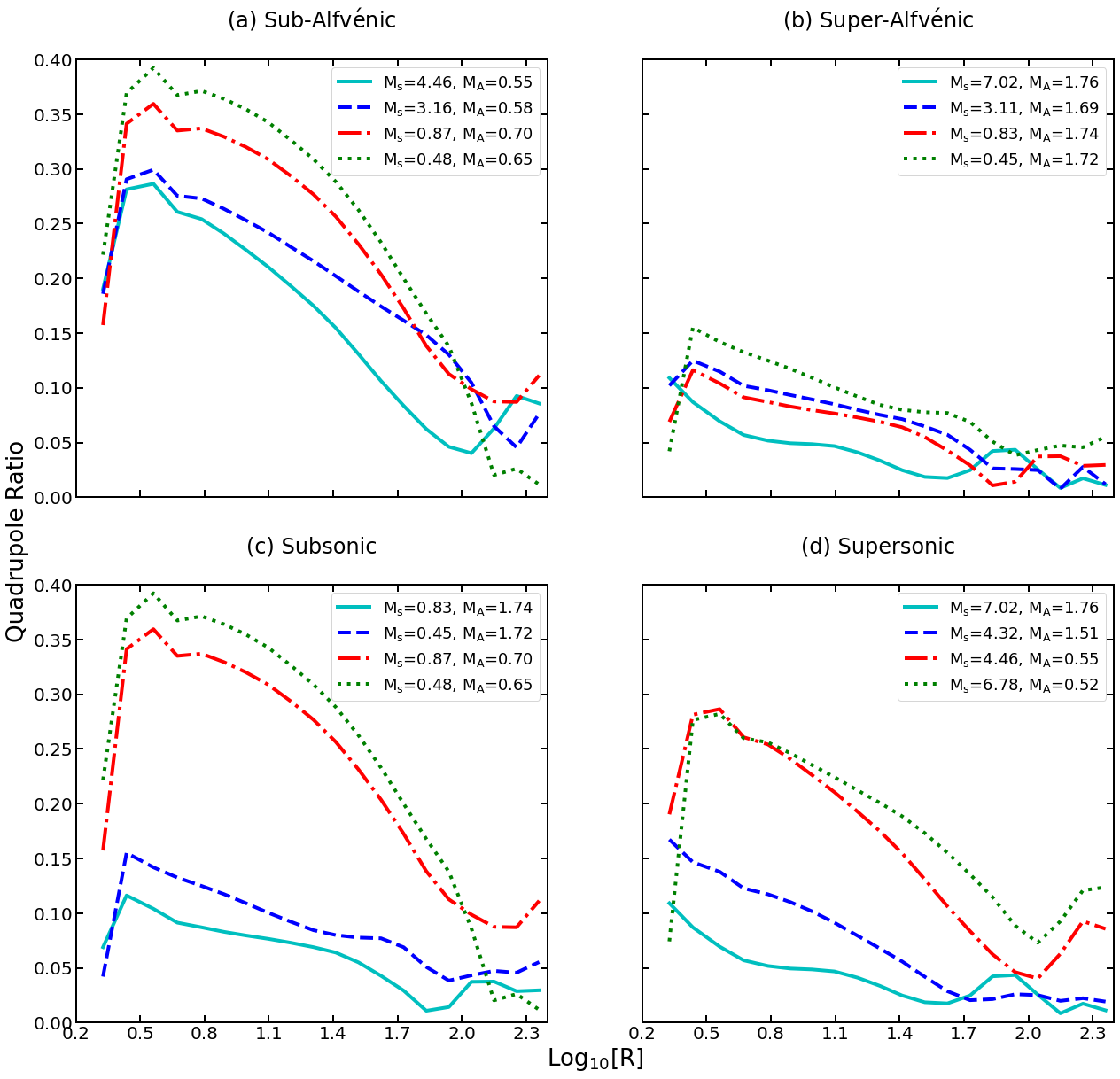}
\caption{Quadrupole ratio modulus of synchrotron polarization intensity as a function of radial separation of maps at different turbulence regimes.
}
\label{Fig:QuadMaMs1}
\end{figure*}

To see the dependence of the anisotropy on Mach numbers more quantitatively, we apply the quadrupole ratio (see Equation \ref{eq:9}) to depict the anisotropy of polarization intensity. Based on the data listed Table \ref{table_1}, quadrupole ratio moduli of polarization intensity, as a function of radial separation, are plotted in Figure \ref{Fig:QuadMaMs1} at four different turbulence regimes. With increasing radial separation, quadrupole ratio moduli decrease and reach minimum value with physical meaning at the scale approximately 100 pixels. Between this scale and the injection scale of $\simeq 200$ pixels, unstable curves reflect the phenomenon that the turbulence has not been fully developed. The curves of quadrupole ratio moduli show a trough at small scales of about 5 pixels, which is due to the existence of the discrete grid of pixels in 2D normalized structure function. 

As for the upper panels of Figure \ref{Fig:QuadMaMs1}, the quadrupole ratio modulus has the largest value of amplitude approximately 0.39 in the left panel, whereas the amplitudes are smaller than 0.15 for a wide range of radial separation in the right panel. It is easy to understand that quadrupole ratio moduli are sensitive to sub-Alfv\'enic Mach numbers, resulting in more anisotropy, while super-Alfv\'enic simulations present a weak anisotropy at small scales and almost isotropy at large scales.\footnote{It should be emphasized that the amplitude of quadrupole ratio modulus is an important basis for judging anisotropy. According to our test using isotropic synthetic data generated in Zhang et al.~(\citeyear{2016ApJ...825..154Z}), we realize that if the mean amplitude $\lesssim 0.08$, the structure of the map can be considered as isotropy.}  As for the lower panels,  the simulations related to sonic Mach numbers are separated into two parts, in which large quadrupole ratio corresponds to low $M_{\rm A}$, however, small quadrupole ratio to high $M_{\rm A}$. The main reason may be that low $M_{\rm A}$ simulations have a stronger magnetic fluctuation and produce more magnetic-field-dominated anisotropy. It can also be seen that the quadrupole ratio moduli depend on sonic Mach numbers. The slope at large scales of the quadrupole ratio moduli for sub-Alfv\'enic and subsonic simulation (left lower panel) is steeper than those for sub-Alfv\'enic and supersonic simulation (right lower panel). The reason should be that increasing sonic Mach number results in the formation of shock waves of fast fluid motions, which maintains smooth changes for anisotropy of turbulence.

\begin{figure*}[ht]
\center
\includegraphics[width=0.80\textwidth]{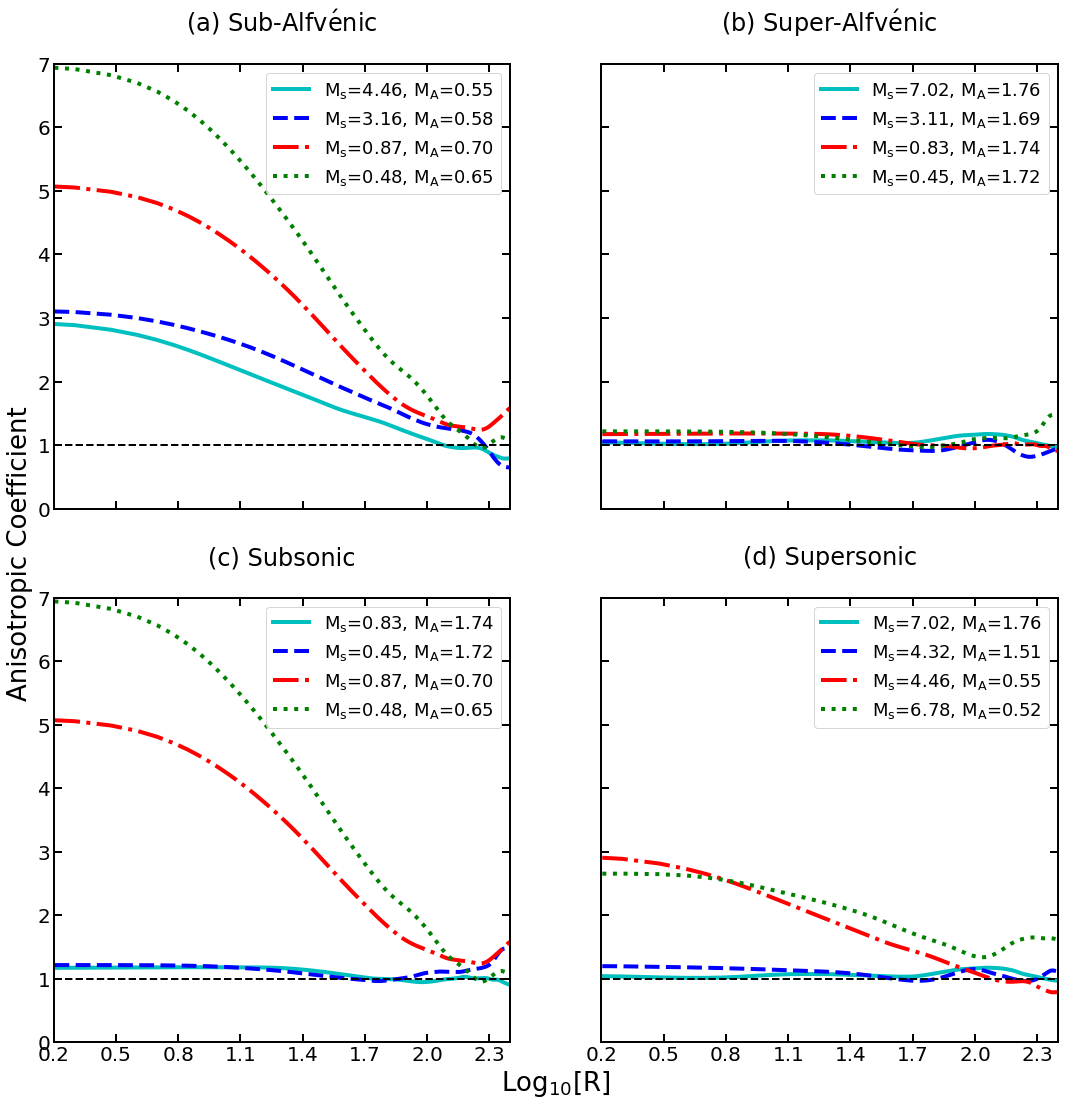}
\caption{Anisotropic coefficient for synchrotron polarization intensity as a function of radial separation of maps at different turbulence regimes. The dashed line plotted in each panel corresponds to anisotropic coefficient of 1 reflecting isotropy properties of MHD turbulence.}
\label{Fig:ACtotal}
\end{figure*}

Alternatively, an additional statistical technique, called anisotropic coefficient, is used to test the anisotropy of synchrotron polarization intensity. This technique can not only reflect the relative value of component of anisotropic structure, but also reveal the direction of mean magnetic field. Anisotropic coefficients of polarization intensity calculated by Equation (\ref{eq:AC}) are plotted in Figure \ref{Fig:ACtotal} using the same data cubes as those of Figure \ref{Fig:QuadMaMs1}.  In each panel of Figure \ref{Fig:ACtotal}, the horizontal dashed line with $AC=1$ represents isotropy of normalized structure function of polarization intensity.
As is seen in Figure \ref{Fig:ACtotal}, anisotropic coefficients of polarization intensity show the scale-dependent anisotropy that decreases as the scale increases. It is clearly shown that the sub-Alfv\'enic simulation generates the anisotropic coefficient values far exceeding those that the super-Alfv\'enic simulation does (weak anisotropy). In other words, anisotropic coefficients are more sensitive to sub-Alfv\'enic Mach numbers, resulting in more significant anisotropy. Besides, we can see that anisotropic coefficients in sub-Alfv\'enic regimes increase as sonic Mach number decreases, which demonstrates the dependence of the anisotropic coefficient on sonic Mach number in sub-Alfv\'enic turbulence regimes. According to the anisotropic coefficient values ($>1$) shown in this figure, we derive the direction of the magnetic field in a horizontal ($x$-axis) direction.

\subsection{Effect of Frequency}\label{EffFreq1}

\begin{figure*}[ht]
\center
\includegraphics[width=0.80\textwidth]{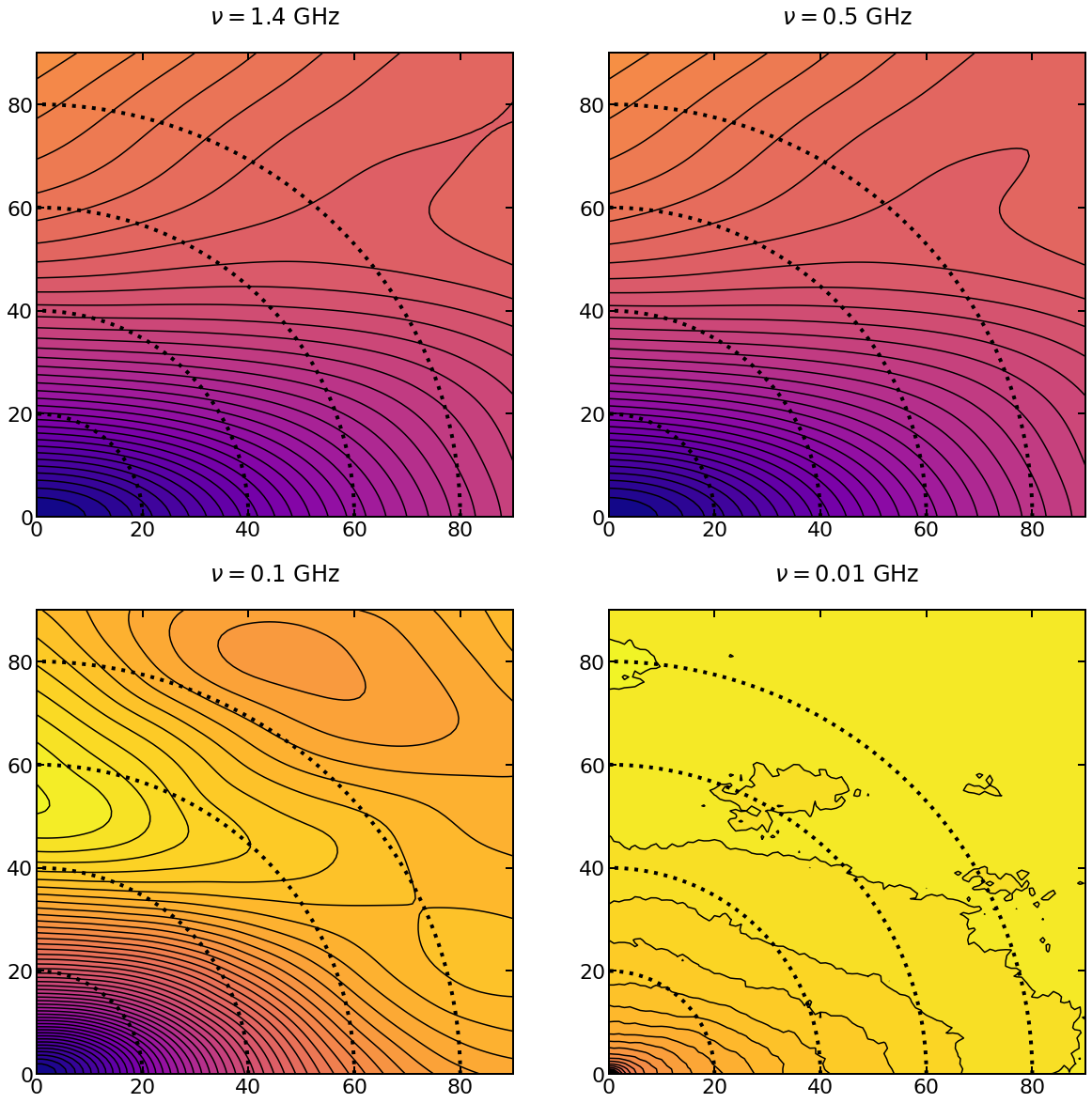}
\caption{Contour maps of normalized structure function of synchrotron polarization intensity at different radiation frequencies. The dotted contour lines plotted in each panel indicate isotropy.
}
\label{Fig:Mapfreq}
\end{figure*}

In the section, we explore how radiative frequency affects the anisotropy of synchrotron polarization intensity using the three methods mentioned above. All calculations are based on the run5 listed in Table \ref{table_1} by setting different frequencies.

Figure \ref{Fig:Mapfreq} shows the contours of normalized structure function of polarization intensity at frequencies of $\nu=$1.4, 0.5, 0.1 and 0.01 GHz. The dotted lines plotted in each panel denote a distribution of the isotropic structure for the sake of comparison. It can be clearly seen that structures of eddies are elongated along $x$-axis, i.e., the direction of mean magnetic field. As the frequency decreases, the elongated anisotropic structures we can see from the figure move to a smaller scale, accompanied by a decrease in anisotropy levels, however, the large-scale anisotropy structure is replaced by the chaotic one. The reason should be that as the frequency decreases, the Faraday rotation produces stronger effect on the anisotropic structure, with the loss of correlation at a large scale. In any case, the Faraday rotation effect will not be an obstacle to studying anisotropy and tracing mean magnetic field directions.

\begin{figure}[ht]
\center
\includegraphics[width=0.48\textwidth]{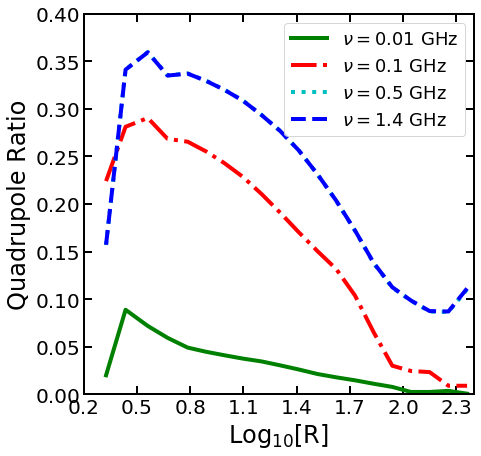}
\caption{
Quadrupole ratio modulus of synchrotron polarization intensity as a function of radial separation at different frequencies. 
}
\label{Fig:QRfreq}
\end{figure}

Next, we use the quadrupole ratio modulus to explore the anisotropy of synchrotron polarization intensity at frequencies of $\nu=$1.4, 0.5, 0.1 and 0.01 GHz. 
The quadrupole ratio modulus of polarization intensity as a function of radial separation is shown in Figure \ref{Fig:QRfreq}. The amplitude of quadrupole ratio modulus decreases with increasing the radial separation, reflecting the anisotropic structures more elongated at small scales. It can be seen that for the simulations plotted at 0.5 and 1.4 GHz, quadrupole ratio moduli present almost the same distributions (changes occurs at about 0.2 GHz according to our test.), however, the amplitude is significantly reduced for the simulation with the frequency of 0.01 GHz. As a result, the low-frequency Faraday rotation causes the anisotropy of polarization intensity to decrease, which will be considered in detail in the future.

\begin{figure}[ht]
\center
\includegraphics[width=0.48\textwidth]{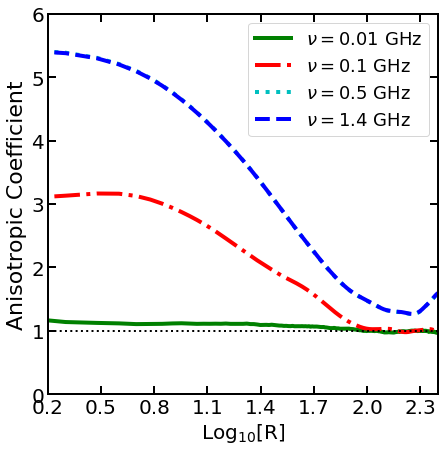}
\caption{Anisotropic coefficient for synchrotron polarization intensity as a function of radial separation of maps at different frequencies. The dotted line indicates isotropy of the map structure. 
}
\label{Fig:ACfreq}
\end{figure}

The influence of the frequency on the anisotropic coefficient is studied in Figure \ref{Fig:ACfreq}, in which the horizontal dotted line represents isotropy. It is shown that the anisotropic coefficient becomes small as the radial separation increases, which reflects the scale-dependent anisotropy. The dependence of the anisotropic coefficient on frequency is similar to those of Figure \ref{Fig:QRfreq}. The difference is that the anisotropic coefficient can directly reveal the direction of mean magnetic field while the quadrupole ratio can reflect anisotropic degree of polarization intensity with high accuracy. Accordingly, combining the quadrupole ratio with the anisotropic coefficient can be very synergetic to study the anisotropy of synchrotron polarization intensity.

%%%%%%%%%%%%%%%%%%%%%%%%%%%%%%%%%%%%%%%%%%%%%%%%%%%%%%
%%                                      Section 5                                       %%%%%%%%%%%%%%%%%%%%%%%
%%%%%%%%%%%%%%%%%%%%%%%%%%%%%%%%%%%%%%%%%%%%%%%%%%%%%%
\section{Anisotropy of polarization intensity for post-decomposition MHD modes}\label{aniThreeMod}

We now explore anisotropies of synchrotron polarized intensities for Alfv\'en, slow and fast modes using quadrupole ratio and anisotropic coefficient methods. As is shown in Section \ref{aniTotMod}, since super-Alfv\'enic simulations show a weak anisotropy, only the sub-Alfv\'enic data listed in Table \ref{table_1} are decomposed in this paper to study the anisotropic structure for the three modes.

\subsection{Influence of Mach Numbers}\label{section4.2.1}

\begin{figure*}[ht]
\centering
\includegraphics[width=0.80\textwidth]{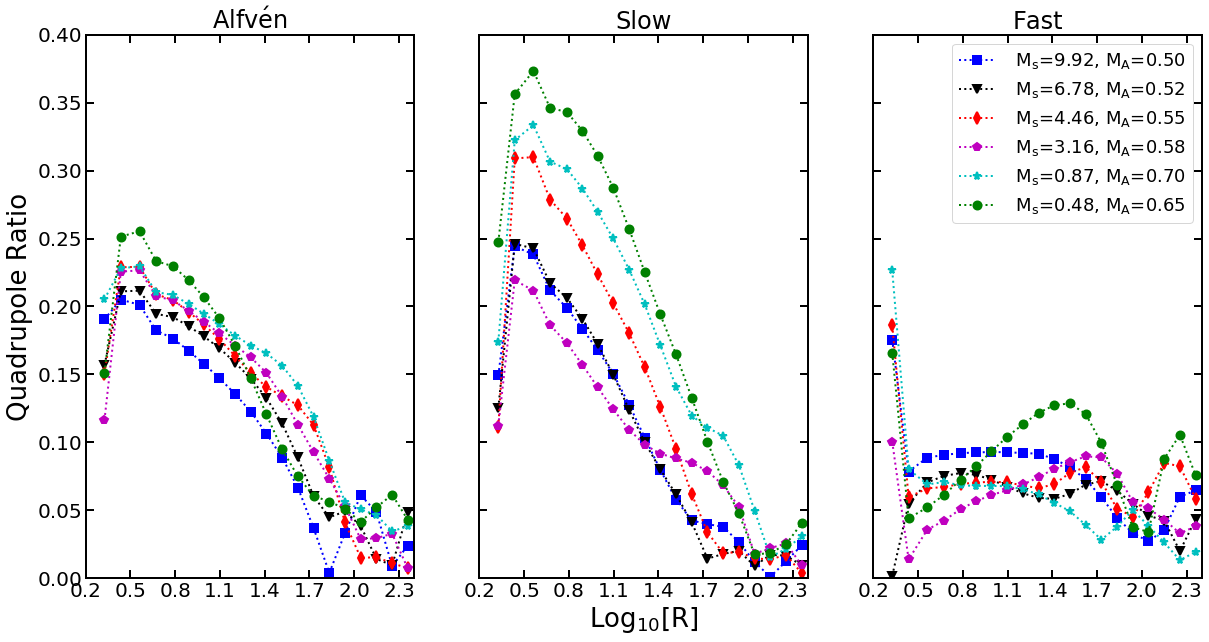}
\caption{Quadrupole ratio moduli of synchrotron polarization intensities for Alfv\'en, slow and fast modes as a function of radial separation of maps in sub-Alfv\'enic regime.
}
\label{Fig:QRMaMs}
\end{figure*}

Using the parameters of electron spectral index of $\alpha=-1$ and the frequency of $\nu=10$ GHz, we show in Figure \ref{Fig:QRMaMs} quadrupole ratio moduli of polarized intensities calculated by Alfv\'en, slow and fast modes. In the left and middle panels of this figure, quadrupole ratio modulus is a decreasing function of radial separation with the peak at approximately 6 pixels, a similar performance to those in the left upper panel of Figure \ref{Fig:QuadMaMs1}, which reveals the scale-related anisotropy of Alfv\'en and slow modes. However, the quadrupole ratio moduli from Alfv\'en modes and slow modes vary in the dependence on $M_{\rm s}$----the former being mild while the latter being sensitive. The reason may be that these two modes have distinct intrinsic characteristics, i.e., incompressible Alfv\'en modes vs. compressible slow modes. Concerning the simulations from slow modes, it can be seen that two curves (marked with $M_{\rm s}=0.48, M_{\rm A}=0.65$ and $M_{\rm s}=0.87, M_{\rm A}=0.70$) present larger amplitudes than others, which suggests that turbulent magnetic pressure dominates small $M_{\rm s}$ fluid pressure to stimulate anisotropic structure. Compared to Alfv\'en modes, most simulations for slow modes have larger amplitude of quadrupole ratio moduli for the same set of simulations. Different from Alfv\'en and slow modes, amplitudes of quadrupole ratio modulus from fast modes (see right panel) have no scale dependencies and meet an anisotropic criterion of the mean amplitude less than 0.08 mentioned above, which means the isotropy of polarized intensity from fast modes.

\begin{figure*}[ht]
\center
\includegraphics[width=0.80\textwidth]{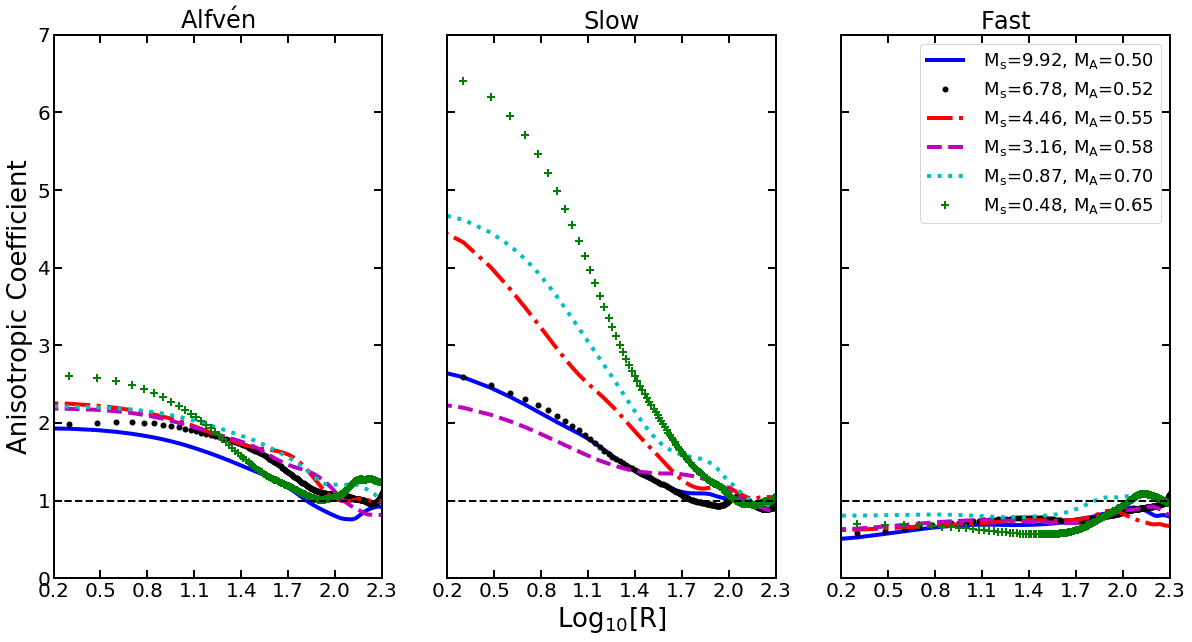}
\caption{Anisotropic coefficients of synchrotron polarization intensity for Alfv\'en, slow and fast modes as a function of radial separation in sub-Alfv\'enic regime. The dashed line plotted in each panel signifies isotropy.
}
\label{Fig:ACThreeMaMs}
\end{figure*}

We now adopt anisotropic coefficient method to explore the degree of anisotropy for three modes, with the results plotted in Figure \ref{Fig:ACThreeMaMs}. It can be clearly seen in the left and middle panels that anisotropic coefficient reflects well scale-dependent anisotropy and its magnitude larger than 1 decreases with the radial separation increasing. As for distributions of the anisotropic coefficient, we find a larger magnitude of the anisotropic coefficient for slow modes than that of Alfv\'en modes in the same set of simulations. On the basis of Equation (\ref{eq:AC}), we know that anisotropic structures of both Alfv\'en and slow modes are elongated along $x$-axis, i.e., the direction of mean magnetic field, which demonstrates that they have a similar anisotropic structure. In the case of the right panel, fast-mode-anisotropic-coefficient magnitude is less than 1 but very close to 1, which indicates that an approximately isotropic structure of polarization intensity should be aligned perpendicular to mean magnetic field. Both Alfv\'en and slow modes dominate the composition of the compressible turbulence; the anisotropic structure of polarized intensity for pre-decomposition MHD modes is confirmed in Section \ref{aniTotMod}.

\subsection{Influence of Spectral Index}\label{SpecIndex}

\begin{figure*}[ht]
\center
\includegraphics[width=0.80\textwidth]{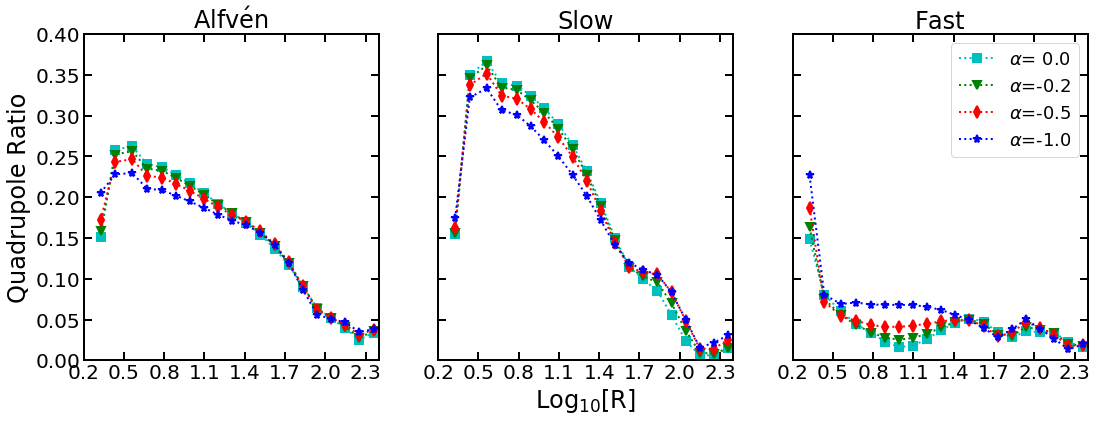}
\caption{Influence of relativistic electron spectral index on quadrupole ratio moduli  of synchrotron polarization intensity for Alfv\'en, slow and fast modes.
}
\label{fig:specindex}
\end{figure*}

In this section, we study the influence of relativistic electron spectral index on the anisotropy of polarized intensity at the fixed frequency of $\nu=10$ GHz, on the basis of the MHD modes decomposition of run5 listed in Table \ref{table_1}.  First, the quadrupole ratio method is used to study the anisotropy of synchrotron polarization intensity, and the results are shown in Figure \ref{fig:specindex}. This figure shows the quadrupole ratio modulus for three modes as a function of radial separation by changing spectral index of relativistic electrons. In general, overall distributions of the quadrupole ratio modulus of polarized intensity are similar to those of Figure \ref{Fig:QRMaMs}. From this figure, we can see that quadrupole ratio moduli for three modes marginally depend on the electron spectral index, the range of which covers the important cases in astrophysics we are aware of.

\begin{figure*}[ht]
\center
\includegraphics[width=0.80\textwidth]{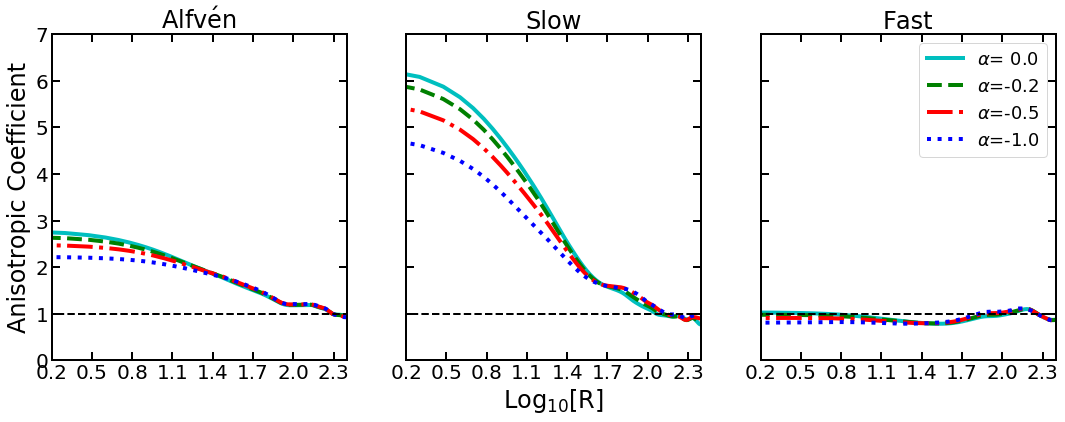}
\caption{Influence of relativistic electron spectral index on anisotropic coefficients of synchrotron polarization intensities for Alfv\'en, slow and fast modes. 
}
\label{Fig:ACthreeMod}
\end{figure*}

Using the anisotropic coefficient method, we now explore the influence of relativistic electron spectral index on the anisotropy of polarized intensities for three modes.  
Figure \ref{Fig:ACthreeMod} shows the relation of anisotropic coefficient and radial separation at different spectral indices, from which we can see that anisotropic coefficient of polarized intensities from slow modes seems more sensitive than that from Alfv\'en modes. In particular, the anisotropic coefficients from fast modes are insensitive to changes in radial separation. As a result, both the quadrupole ratio moduli and anisotropic coefficients from synchrotron polarization intensities marginally depend on the spectral index of relativistic electrons, the change of which does not prevent us from determining the anisotropy of magnetic turbulence.

\subsection{Influence of Frequency}\label{influoffreq}

\begin{figure*}[ht]
\centering
\includegraphics[width=0.80\textwidth]{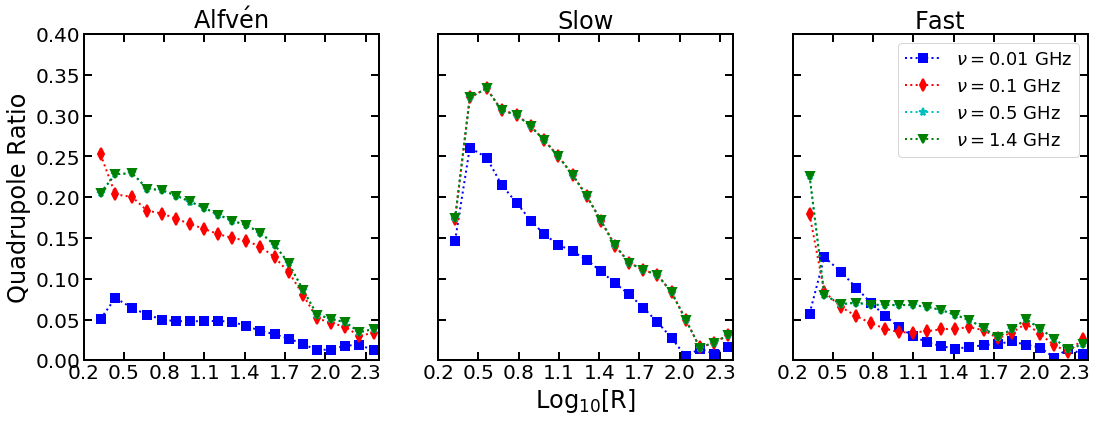}
\caption{Influence of radiation frequency on quadrupole ratio moduli of synchrotron polarization intensity for Alfv\'en, slow and fast modes. }
\label{Fig:QRfreqMod}
\end{figure*}

The influence of the radiative frequency on the quadrupole ratio modulus of polarized intensities for three modes are explored in this section by using electron spectral index of $\alpha=-1$. Based on the MHD-mode-decomposition of run5 listed in Table \ref{table_1}, we adopt two statistical techniques mentioned above, i.e., quadrupole ratio modulus and anisotropic coefficient methods, to reveal the anisotropy of MHD turbulence. Quadrupole ratio moduli for three modes as a function of radial separation are shown in Figure \ref{Fig:QRfreqMod} at different frequencies. From this figure, we find that the quadrupole ratio moduli for three modes marginally depend on the frequency, while the quadrupole ratio modulus for slow modes has smaller dependence on the frequency compared with those of other modes. In general, with decreasing the frequency, we expect the amplitude of quadrupole ratio modulus to become smaller, because the Faraday rotation effect becomes stronger at low frequencies. However, for the frequency ranges we study in this paper, the quadrupole ratio modulus for slow mode does not change much. Therefore, quadrupole ratio modulus with larger amplitude from slow mode provides a more reliable way to determine anisotropy of magnetic turbulence than those of Alfv\'en and fast modes.

\begin{figure*}[ht]
\center
\includegraphics[width=0.80\textwidth]{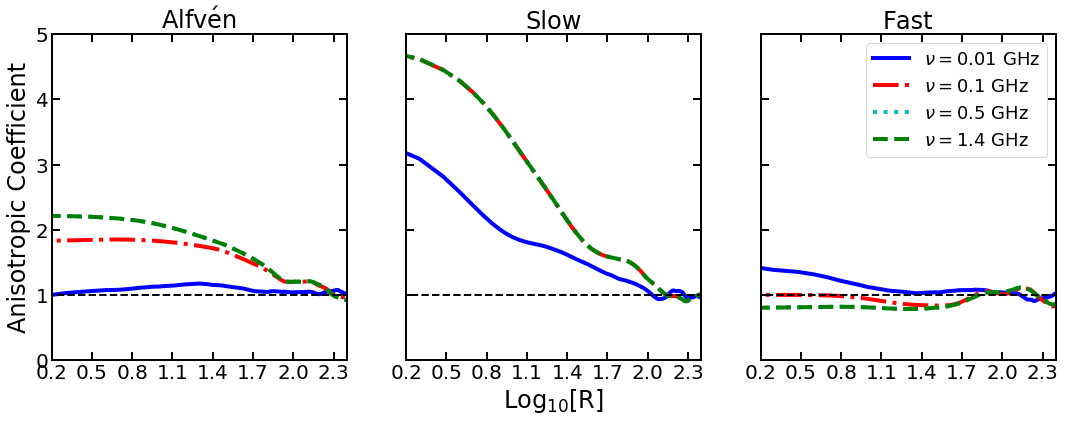}
\caption{Influence of radiation frequency on anisotropic coefficients of synchrotron polarization intensity for Alfv\'en, slow and fast modes.
}
\label{Fig:ACfreqMod}
\end{figure*}

Exploring anisotropy of polarization intensities for three modes at different frequencies is plotted in Figure \ref{Fig:ACfreqMod} by the anisotropic coefficient method. 
As is shown in this figure, anisotropic coefficients of polarized intensities for three modes show a behavior similar to those of Figure \ref{Fig:QRfreqMod}. Namely, anisotropic coefficients for slow mode are less sensitive to the frequency while anisotropic coefficients for Alfv\'en and fast modes depend on the frequency. As a result, the low-frequency strong Faraday rotation effect slightly reduces the anisotropic degree of polarization intensity, whose structures are elongated along mean magnetic field direction for Alfv\'en and slow modes.

%%%%%%%%%%%%%%%%%%%%%%%%%%%%%%%%%%%%%%%%%%%%%%%%%%%%%%
%%                                      Section 6                                       %%%%%%%%%%%%%%%%%%%%%%%
%%%%%%%%%%%%%%%%%%%%%%%%%%%%%%%%%%%%%%%%%%%%%%%%%%%%%%

\section{Discussion}\label{Discus}

This work studied the anisotropy of synchrotron polarized intensity using 2D normalized structure function, quadrupole ratio modulus and anisotropic coefficient. The contour map of the structure function can qualitatively reveal the anisotropy of MHD turbulence, while anisotropic coefficient adopting the ratio of parallel component of the structure function to perpendicular one can quantitatively measure the anisotropy. Importantly, the quadrupole ratio modulus can accurately reflect the degree of anisotropy at different scales. 

The structure function map can be used as a prior for diagnosing the anisotropy of synchrotron polarization intensity. The synergy effect can be realized in the quantitative study of the anisotropy of synchrotron polarization intensity by the combination of the quadrupole ratio and the anisotropic coefficient. The former can be used to accurately measure the amplitude of anisotropy with no reflection of mean magnetic field direction. The latter reveals the degree of the anisotropy by the ratio of the parallel to perpendicular component of the structure function with the successful reflection of mean magnetic field direction. Provided with the anisotropic coefficient greater than 1, the parallel component of synchrotron polarization intensity is larger than the perpendicular one, and vice versa. 

Based on multiple sets of data, we first tested in Section \ref{EffFreq1} the results Lee et al. (\citeyear{2019ApJ...877..108L}) got when they studied how the wavelength affects the anisotropy of synchrotron polarized intensity. Then, we move to focus on the investigation of the anisotropy of MHD turbulence. Generally, the structures of polarized intensities from Alfv\'en and slow modes are found to be anisotropic, being aligned with the direction of the mean magnetic field, while those from fast modes are isotropic, with the tendency of being perpendicular to the direction of the mean magnetic field. The results from the above three modes are consistent with that from the direct numerical simulation of CL02 and CL03, only with a slightly stronger anisotropy of the slow modes than that of Alfv\'en modes. Meanwhile, the slow mode simulation shows less dependence on the radiation frequency. 

Therefore, more robust measurements of magnetic field can be achieved by means of the synchrotron polarization statistics which restored the MHD turbulence modes successfully. The distinct properties of each mode revealed in our simulation provide us the possibility to distinguish the different contributions which they make in the real astrophysical environment. Complementary to the synchrotron gradients technique mentioned previous to map 3D structure of the magnetic field, the application of the second-order normalized structure function, quadrupole ratio modulus and anisotropic coefficient provides a more detailed analysis for the contribution of the compressible MHD turbulence modes.

As is shown in Figures \ref{Fig:QRfreq}, \ref{Fig:ACfreq}, \ref{Fig:QRfreqMod} and \ref{Fig:ACfreqMod}, the anisotropic level of polarization intensity is relatively low at the frequency of 0.01 GHz, which makes the low-frequency polarized intensity more challenging to reveal the properties of MHD turbulence in the effect of the relevant noise and strong Faraday rotation. If the effect of small-scale noise-like is removed, we expect an increase in the degree of anisotropy, which should be similar to the scenario for the application of synchrotron polarization gradient techniques in the low frequency range (Zhang et al. \citeyear{2019MNRAS.486.4813Z,2019arXiv191002378Z}). The quadrupole ratio modulus and anisotropic coefficient used in this paper are complementary techniques to the gradient measurement. With the high-resolution data presently available from Low Frequency Array for radio astronomy and those in the future from the Square Kilometer Array, these techniques will have broader application prospects.

%%%%%%%%%%%%%%%%%%%%%%%%%%%%%%%%%%%%%%%%%%%%%%%%%%%%%%
%%                                      Section 7                                       %%%%%%%%%%%%%%%%%%%%%%%
%%%%%%%%%%%%%%%%%%%%%%%%%%%%%%%%%%%%%%%%%%%%%%%%%%%%%%

\section{Summary}\label{Summ}
The polarized intensity anisotropy has been investigated in this paper, using statistical techniques of normalized structure function, quadrupole ratio and anisotropic coefficient to reveal the anisotropy of compressible MHD turbulence. We focused on studying how Mach numbers, spectral index and radiation frequency affect the anisotropy of polarized intensity and exploring how to obtain the anisotropy of eddies related to the direction of mean magnetic field. The main results are briefly summarized as follows.

\begin{enumerate}[wide, labelwidth=!, labelindent=1pt]
\item The anisotropy of synchrotron polarization intensity is found to be influenced by several MHD turbulence regimes. The most significant anisotropy of polarized intensity occurs in the sub-Alfv\'enic turbulence regime, in which a gradual change in anisotropy happens in the supersonic regime with the increase of the spatial scale. 

\item With the decreasing frequency, the level of the anisotropy of polarization intensity is reduced due to the Faraday rotation effect. But that will not be an obstacle to studying MHD turbulence anisotropy and tracing mean magnetic field directions.

\item On the basis of basic MHD modes decomposed using the sub-Alfv\'enic data cubes, we have found that\\
(a) the anisotropy of the polarized intensity structure from Alfv\'en modes can help better trace the direction of the mean magnetic field, with slight reliance on the electron spectral index and the radiation frequency.\\
(b) the polarized intensity structure from slow modes is also anisotropic, showing a stronger anisotropy compared to that from Alfv\'en modes.  
The anisotropy for slow modes has marginal dependence on electron spectral index and frequency.\\
(c) the polarized intensity structure from fast modes is nearly isotropic. For fast modes, the influence of electron spectral index and frequency on the structure of polarized intensity is insignificant. \\

\item The elongated structures of the polarized intensity from Alfv\'en and slow modes are aligned with the direction of the magnetic field. These findings are in good 
consistency with those from the earlier direct numerical simulations.

\item The normalized structure function, anisotropic coefficient and quadrupole ratio modulus are very synergetic in studying the anisotropy of compressible MHD turbulence. The former two can reflect the direction of mean magnetic field and the latter one can accurately obtain the degree of anisotropy.

\end{enumerate}

\acknowledgments
We thank the anonymous referee for valuable comments, Alex Lazarian for useful discussion and Christopher Herron for providing python code to calculate quadrupole ratio modulus. J.F.Z. thanks the supports from the National Natural Science Foundation of China (grant Nos. 11973035 and 11703020) and the Hunan Provincial Natural Science Foundation (grant No. 2018JJ3484). F.Y.X. acknowledges the support of  the National Natural Science Foundation of China (grant No. U1731106). R.Y.W. and J.F.Z. thank the support of Guizhou Provincial Key Laboratory of Radio Astronomy and Data Processing (grant No. KF201803).


\begin{thebibliography}{}

\bibitem[Armstrong et al.(1995)]{1995ApJ...443..209A} Armstrong, J.~W., Rickett, B.~J., \& Spangler, S.~R.\ 1995, \apj, 443, 209
\bibitem[Beck \& Wielebinski (2013)]{2013pss5.book..641B} Beck, R., \& Wielebinski, R.\ 2013, in Planets, Stars and Stellar Systems, Vol. 5, 
ed. T. D. Oswalt \& G. Gilmore (Dordrecht: Springer), 641
\bibitem[Brentjens \& de Bruyn (2005)]{2005A&A...441.1217B} Brentjens, M., \& de Bruyn, A.\ 2005, \aap, 441, 1217
\bibitem[Burn (1966)]{1966MNRAS.133...67B} Burn, B.~J.\ 1966, \mnras, 133, 67
\bibitem[Burkhart et al. (2012)]{2012ApJ...749..145B} Burkhart, B., Lazarian, A., \& Gaensler, B.~M.\ 2012, \apj, 749, 145
\bibitem[Cho \& Lazarian (2002)]{2002PhRvL..88x5001C} Cho, J., \& Lazarian, A.\ 2002, PhRvL, 88, 245001 
\bibitem[Cho et al. (2002)]{cho2002simulations} Cho, J., Lazarian, A, \& Vishniac, E. T.\ 2002, \apj, 564, 291
\bibitem[Cho \& Lazarian (2003)]{2003MNRAS.345..325C} Cho, J., \& Lazarian, A.\ 2003, \mnras, 345, 325
\bibitem[Cho \& Vishniac (2000)]{2000ApJ...539..273C} Cho, J., \& Vishniac, E.~T.\ 2000, \apj, 539, 273
\bibitem[Dickey et al. (2019)]{2019ApJ...871..106D} Dickey, J. M., Landecker, T.~L., Thomson, A.~J.~M., et al.\ 2019, \apj, 871, 106
\bibitem[Elmegreen \& Scalo (2004)]{2004ARA&A..42..211E} Elmegreen, B.~G., \& Scalo, J.\ 2004, \araa, 42, 211 
\bibitem[Ferri{\`e}re (2001)]{2001RvMP...73.1031F} Ferri{\`e}re, K.~M.\ 2001, RvMP, 73, 1031
\bibitem[Fletcher et al. (2011)]{2011MNRAS.412.2396F} Fletcher, A., Beck, R., Shukurov, A., et al.\ 2011, \mnras, 412, 2396
\bibitem[Gaensler et al. (2011)]{2011Natur.478..214G} Gaensler, B.~M., Haverkorn, M., Burkhart, B., et al.\ 2011, Nature, 478, 214
\bibitem[Ginzburg \& Syrovatskii (1965)]{1965ARA&A...3..297G} Ginzburg, V.~L., \& Syrovatskii, S.~I.\ 1965, \araa, 3, 297 
\bibitem[Goldreich \& Sridhar (1995)]{1995ApJ...438..763G} Goldreich, P., \& Sridhar, S.\ 1995, \apj, 438, 763 
\bibitem[Hamaker \& Bregman (1996)]{1996A&AS..117..161H} Hamaker, J.~P., \& Bregman, J.~D.\ 1996, \aaps, 117, 161
\bibitem[Haverkorn (2015)]{2015ASSL..407..483H} Haverkorn, M.\ 2015, ASSL, 407, 483
\bibitem[Herron et al. (2016)]{Herron16} Herron, C. A., Burkhart, B., Lazarian, A., et al.\ 2016, \apj, 822, 13
\bibitem[Herron et al. (2018b)]{2018bApJ...853....9H} Herron, C.~A., Gaensler, B.~M., Lewis, G.~F., et al.\ 2018b, \apj , 853, 9 
\bibitem[Iroshnikov (1964)]{1964SvA.....7..566I} Iroshnikov, P.~S.\ 1964, SvA, 7, 566
\bibitem[Jeli{\'c} et al. (2015)]{2015A&A...583A.137J} Jeli{\'c}, V., de Bruyn, A.~G., Pandey, V.~N., et al.\ 2015, \aap, 583, A137
\bibitem[Kraichnan (1965)]{1965PhFl....8.1385K} Kraichnan, R.~H.\ 1965, PhFl, 8, 1385
\bibitem[Kowal \& Lazarian (2010)]{2010ApJ...720..742K} Kowal, G., \& Lazarian, A.\ 2010, \apj, 720, 742
\bibitem[Lazarian (2006)]{2006ApJ...645L..25L} Lazarian, A.\ 2006, \apjl, 645, L25
\bibitem[Lazarian \& Pogosyan (2012)]{2012ApJ...747....5L} Lazarian, A., \& Pogosyan, D.\ 2012, \apj, 747, 5
\bibitem[Lazarian \& Pogosyan (2016)]{2016ApJ...818..178L} Lazarian, A., \& Pogosyan, D.\ 2016, \apj, 818, 178
\bibitem[Lazarian \& Vishniac (1999)]{1999ApJ...517..700L} Lazarian, A., \& Vishniac, E.~T.\ 1999, \apj, 517, 700 
\bibitem[Lazarian \& Yuen (2018)]{2018bApJ...865...59L} Lazarian, A., \& Yuen, K.~H.\ 2018, \apj, 865, 59
\bibitem[Lee et al. (2019)]{2019ApJ...877..108L} Lee, H., Cho, J., \& Lazarian, A.\ 2019, \apj, 877, 108
\bibitem[Lithwick \& Goldreich (2001)]{Lithwick01} Lithwick, Y., \& Goldreich, P.\ 2001, ApJ, 562, 279
\bibitem[Mac Low \& Klessen (2004)]{2004RvMP...76..125M} Mac Low, M.-M., \& Klessen, R. S.\ 2004, RvMP, 76, 125
\bibitem[Maron \& Goldreich (2001)]{Maron01} Maron, J., \& Goldreich, P.\ 2001, ApJ, 554, 1175
\bibitem[Matthaeus et al. (1996)]{1996JGR...101.7619M} Matthaeus, W. H., Ghosh, S., Oughton, S., Roberts, D. A.\ 1996, JGR, 101, 7619
\bibitem[McKee \& Ostriker (2007)]{2007ARA&A..45..565M} McKee, C.~F., \& Ostriker, E.~C.\ 2007, \araa, 45, 565
\bibitem[Narayan \& Medvedev (2001)]{2001ApJ...562L.129N} Narayan, R., \&  Medvedev, M.~V.\ 2001, \apjl, 562, L129 
\bibitem[Shebalin et al. (1983)]{1983JPlPh..29..525S} Shebalin, J.~V., Matthaeus, W.~H., \& Montgomery, D.\ 1983, JPlPh, 29, 525
\bibitem[Van Eck et al. (2017)]{2017A&A...597A..98V} Van Eck, C.~L., Haverkorn, M., Alves, M.~I.~R., et al.\ 2017, \aap, 597, A98
\bibitem[Yan \& Lazarian (2008)]{2008ApJ...673..942Y} Yan, H., \& Lazarian, A.\ 2008, \apj, 673, 942
\bibitem[Zhang et al. (2016)]{2016ApJ...825..154Z} Zhang, J.-F., Lazarian, A., Lee, H., \& Cho, J.\ 2016, \apj, 825, 154
\bibitem[Zhang et al. (2018)]{2018ApJ...863..197Z} Zhang, J.-F., Lazarian, A., \& Xiang, F.-Y.\ 2018, \apj, 863, 197
\bibitem[Zhang et al. (2019a)]{2019MNRAS.486.4813Z} Zhang, J.-F., Lazarian, A., Ho, K.~W., et al.\ 2019a, \mnras, 486, 4813
\bibitem[Zhang et al. (2019b)]{2019arXiv191002378Z} Zhang, J.-F., Liu, Q., \& Lazarian, A.\ 2019b, ApJ, 886, 63


\end{thebibliography}
\end{document}